\documentclass[10pt]{article}          

\usepackage{graphicx}
\usepackage[dvips]{color}
\newcommand\Softitle[1]{\Large \bf \noindent \begin{center} #1
\end{center}\rm \normalsize \vskip.125in }%

\newcommand\Sofauthor[1]{\vskip.1in\noindent%
   \large  \begin{center} \textsf{#1} \end{center}\rm \vskip-.2in}

\long\def\symbolfootnote[#1]#2{\begingroup%
\def\thefootnote{\fnsymbol{footnote}}\footnote[#1]{#2}\endgroup}

\let\title\Softitle
\let\author\Sofauthor

\let\address\Sofaddress
\let\email\Sofemail

\begin{document}

\title{Corpuscular description of the speed of light in a homogeneous medium}

\author{Marcel URBAN, Fran\c cois COUCHOT, Sylvie DAGORET-CAMPAGNE and Xavier SARAZIN}
\address{LAL, Univ Paris-Sud, IN2P3/CNRS, Orsay, France}
\email{urban@lal.in2p3.fr}

\begin{abstract}
We are used to describe the detection of light in terms of particles and its propagation from the source to the detection, by waves.  For instance, the slowing down of light in a transparent medium is always explained within the electromagnetic wave framework. We propose to approach that phenomenon through a purely corpuscular description. We find expression for the refractive indices which differ slightly from the usual Maxwell wave approach. We thus compare these expressions against experimental refractive indices and we show that both reproduce well the data. We show also how this corpuscular framework gives a very natural interpretation to the self focusing Kerr effect. Finally an experimental expectation of fluctuation of the speed of light is presented.

\end{abstract}

\section{Introduction}
\label{sec:introduction}

We will be, here, interested in one well known phenomenon unveiled by Leon Foucault. He showed experimentally \cite{foucault-1850} that light slows down in water with respect to vacuum. More generally, visible light velocity is reduced in transparent media. In a homogeneous material this is described by the index of refraction $n$, which is the ratio of the velocity of light in vacuum to its velocity in the medium.

The refraction phenomenon consists of two observed facts: the bending of light at the interface between two transparent media and the velocity of light being reduced. 

The breaking of the light rays is, most of the time, explained through wave interferences but a particle point of view is possible also. It requires the fact that a single photon is either reflected or transmitted, and never both \cite{grangier-1986}. The energy of the photon is conserved either when reflected or when transmitted, the momentum of the photon is larger in a medium than in vacuum by the factor $n$ \cite{ashkin-1973}\cite{campbell-2005}, and the projection of the momentum onto the boundary is conserved because of translational invariance along that direction. 

To the best of our knowledge there is only one explanation to the second fact of the smaller velocity of light and it is a wave framework (see appendix A).

We describe, in this paper, a particle approach of the slowing down of light in a transparent medium.

We describe our model in section~\ref{sec:corpuscular}. 
Then we show in section~\ref{sec:comparison} that the index of refraction calculated by our corpuscular expression is in good agreement both with the usual Maxwell-quantum calculation and with a few experimental data.

With the advent of lasers in the $60$'s it was found that focusing intense laser beams on glass would produce filaments several centimetres long and a few wavelengths in diameter. The paradox is that diffraction should spread the beam over a distance of a few $10^{-4}$cm. The corpuscular approach interprets naturally this self focusing Kerr phenomenon and we derive its magnitude in gases and in condensed matter in section~\ref{sec:kerr}. 

Finally an experimental expectation of fluctuations of the speed of light is presented in the section~\ref{sec:flower}. It is shown that its measurement could be within the range of today's technical capabilities.

\section{A corpuscular approach}
\label{sec:corpuscular}

It is an experimental fact that molecules have discrete energy states, and in a gas these states are very well separated as opposed to solids where they can merge and form continuous bands. One problem with a corpuscular description of light in propagation is that a photon is absorbed by a molecule only if its energy corresponds to an allowed difference of energies in this molecule. For instance if the first excited state is  $10 \ eV$ above the ground state, visible photons of a few $eV$ should do nothing to these molecules. Why then do we have an index of refraction for a continuum of photon energies? Or, put differently, how come a photon whose energy does not fit an atomic line can be influenced by the medium?

From the point of view of quantum mechanics, energy non-conservation is a natural phenomenon for short durations. So, the system photon-molecule can be thought of as borrowing or putting aside for a while the amount of energy allowing it to bring the molecule to an excited state. This virtual excited state absorbs the initial photon momentum $p=E/c$. As the molecular momentum at room temperature is a few $keV/c$ while the photon momentum is a few $eV/c$, the molecule continues, almost undisturbed, at its thermal speed. The virtual excited state propagates at speeds negligible compared to the speed of light because the thermal speed is, even for atomic hydrogen, a few $10^3 m/s$. Therefore the virtual state is almost equivalent to a stop time for the photon.

After a while, the molecule and the photon are brought back to their initial energy and momentum states. This way, light goes on straight, having only undergone a small time delay in this interaction with the molecule. From a thermo-dynamical point of view, this phenomenon is reversible and does not increase entropy. It looks like decay stimulated by the incoming photon responsible for bringing the molecule to the excited state. 

When the photon has just the amount of energy needed to put the molecule in one of its excited states,
it can be absorbed. In this case, the excited state is real. It decays spontaneously with the lifetime associated to the state. This effect is not reversible. It creates entropy. Light is absorbed and reemitted isotropically. This phenomenon dominates the behaviour of the medium when photon energies lie in the absorption bands of the medium.

\subsection{Propagation through a gas}
\label{sec:propagas}
The photon has an energy $E_{\gamma}$ and is propagating through molecules in their ground state of energy $W_0$. 
The molecules have an infinite number of discrete excited energy states $W_i$.  In order for the real 
photon to excite the molecule by going from its ground state to the excited level of energy $W_i$, an 
amount $\Delta E_i = W_i - W_0 - E_{\gamma}$  of energy has to appear for a duration $\tau_i$.

$$\tau_i = K_1 {\hbar / {\Delta E_i}}$$

We have introduced a constant $K_1$ which value should be around 1.
After that time the molecule returns to its ground state, the photon reappears and resumes its trip with the vacuum velocity and in the exact same direction. The only observable effect is a shift in time (and in space but at the thermal speed of $\beta=10^{-6}$, this spatial shift is only a few $10^{-15}$m). 
The elastic or Rayleigh scattering is present but this occurs at a very low rate. For instance a visible 
photon has to cross $100$ km of air before having such an elastic interaction.  

The stop time is defined as the average of the $\tau_i$:  

$$t_{stop} = < \tau_i > = K_1 \hbar {\sum_{i}{{f_i}\over {W_i-W_0-E_{\gamma}}}}$$

where $f_i$ is the relative strength of the transition $W_0 \to W_i$.

We simplify this formula by defining an average molecular excitation energy $E_{AV}$ such that:
$$\sum_{i}{{f_i}\over {W_i-W_0-E_{\gamma}}}\approx{{1}\over {E_{AV}-E_{\gamma}}}$$
Here we symplify the expression to a single resonance. We will show in section~\ref{sec:comparison} that it is enough to reproduce well the experimental data in a limited energy range. As in the wave approach expression, two or three resonances may be required to fit the data in a larger energy range in infra red.

The stop time is then given by:
\begin{eqnarray}
\label{eqtstopun}
t_{stop} = K_1 {{\hbar}\over {E_{AV}-E_{\gamma}}}
\end{eqnarray}

As it will be described in section~\ref{sec:airdata}, the average molecular excitation energy $E_{AV}$ in air derived from the refraction index measurements is $E_{AV} = 19\ eV$. Thus the stop time for photons in visible range is about $5\ 10^{-17}s$.
It is way shorter than the life time of an excited level which stands in the nanosecond range, but it is larger than the time for light to cross an atom which is about $3.10^{-19}$s.

Let the mean free path between collisions be $\Lambda$. The total average time to cross a distance
$\Lambda$ is $t_{stop}+\Lambda/c$, and the average speed $V$ is:

 $$V = {\Lambda \over{t_{stop}+\frac{\Lambda}{c}}}$$
 
 Finally, the refractivity is

\begin{eqnarray}
\label{eqtstopdeux}
n-1 ={c\over V} -1={c\over \Lambda}\left({t_{stop}+{\Lambda\over c}}\right) -1 = {c t_{stop}\over \Lambda}
\end{eqnarray}

When the energy of the incoming real photon is increased, the energy violation is smaller and, consequently, the stop time is longer. This explains the dispersion of the index of refraction $n$, with the photon energy $E_{\gamma}$: $dn / dE_{\gamma} > 0$.

The mean free path for a photon is given by:
$$\Lambda = {1\over{\sigma N_{mol}}}$$

where $N_{mol}$ is the number of molecules per unit volume and $\sigma$ is the cross section for a real photon to excite momentarily a molecule.

There are no experimental data about this process except the index of refraction. We assume this cross section to be close to the product of the transverse geometric cross section  $\sigma_{\bot}$ of the molecule and of $\alpha$, the fine structure constant characteristic of the efficiency of photon interactions. Thus we can write:
\begin{eqnarray}
\label{eqsigma}
\sigma = K_2 \alpha \sigma_{\bot}
\end{eqnarray}

In appendix B we derive an estimate of $K_2$ in the framework of quantum mechanic and perturbative theory and it turns out to be of order unity.

The mean free path for a photon is then:
$$\Lambda = {1\over{K_2\alpha\sigma_{\bot} N_{mol}}}$$

Finally our prediction for the refractivity is:

\begin{eqnarray}
\label{eqcorp}
\left[{n-1}\right]_{corpuscle}={K_2\alpha\sigma_{\bot} N_{mol}}\ {{K_1 \hbar c}\over {E_{AV}-E_{\gamma}}}=K \alpha \sigma_\bot N_{mol}  \left( \frac{\hbar c}{E_{AV}-E_\gamma}\right)
\end{eqnarray}

with $K=K_1 K_2$.

$\sigma_{\bot}$ is estimated from the mass of a mole $M$, the Avogadro number ${\cal N}_A$ and the density $\rho$ of the liquid or solid state when they exist.
$$\sigma_{\bot}=\frac{\pi}{4}\left( \frac{M}{{\cal N}_A\rho}\right)^{\frac{2}{3}}$$

\subsection{Propagation through a condensed medium}
\label{sec:propacondensed}
The formula for the index of refraction in a condensed medium is going to be simplified with respect to what we have in a gas. In a crystal we have the molecules packed and touching each other. The numerical density of these molecules can be expressed in terms of the spacing between them. 
If we have three axes $Ox,Oy$ and $Oz$ and the corresponding spacing $\delta_x, \delta_y$ and $\delta_z$, then:

$$N_{mol}=\frac{1}{\delta_x\delta_y\delta_z}$$
If the photon is propagating along Ox we have:
$$\sigma{_\bot}=\delta_y\delta_z\ \  \Rightarrow\ \  {K\alpha\sigma_{\bot} N_{mol}}=K{\alpha\over\delta_x}$$

$$\left[{n_x-1}\right]_{corpuscle}={K\alpha\sigma_{\bot} N_{mol}}\ {{\hbar c}\over {E_{AV}-E_{\gamma}}}=K{\alpha\over\delta_x}{{\hbar c}\over {E_{AV}-E_{\gamma}}}$$
This shows that having a cross section in our index formula leads to sensitivity to the spacing in the direction of propagation. When the molecule is very asymmetrical and fixed in position like in a crystal, the index depends upon the direction of propagation of the light in the crystal.

\section{Comparing the corpuscular and the Maxwell wave expressions to the experimental data}
\label{sec:comparison}

\subsection{The standard wave approach: the Maxwell-quantum formula}

All measurements of refraction indices are summarized through either one of the old empirical formulas:
$$n_{Cauchy}=A+B{\lambda}^2+C{\lambda}^4\ (1840),$$
$$n_{Sellmeier}=\sqrt{1+\sum_{i=1}^3{\frac{a_i^2{\lambda}^2}{{\lambda}^2-b_i^2}}} \ (1871),$$
$${\rm or\ } n_{Hartmann}=A+\frac{B}{\lambda-\lambda_0}{\rm , by}\ 1900.$$
Most of the time, two terms are enough to reproduce the data. 


The standard wave approach described in appendix A leads to the so called Maxwell-quantum formula of the index (A.1).
Since the theoretical Maxwell-quantum index is obtained through the phase shift of the incident plane wave, it corresponds to the phase index $n_p$.
As we did for the corpuscular model we symplify the expression here to a single resonance and we approximate the formula~(A.1) by a single average excitation energy $E_{AV}$:
\begin{eqnarray}
\label{eqmax}
\left[{n_p-1}\right]_{Maxwell}=N_{elec}2\pi r_e\ {{(\hbar c)}^2\over {E_{AV}^2-E_{\gamma}^2}}
\end{eqnarray}
The number density $N_{elec}$ of dispersion electrons is given by
$$ N_{elec} =  n_{valence} N_{mol} $$
where $N_{mol}$ is the number density of molecules and $n_{valence}$ is the number of valence electron corresponding to the number of electrons missing to get a complete shell (for instance in $N_2$, $n_{valence} = 2\times3 = 6$). The equation~\ref{eqmax} becomes

\begin{eqnarray}
\label{eqmaxval}
\left[{n_p-1}\right]_{Maxwell}=n_{valence}2\pi r_e\ N_{mol}{{(\hbar c)}^2\over {E_{AV}^2-E_{\gamma}^2}}
\end{eqnarray}

One problem with the light wave or Maxwell-quantum framework is the fact that there is no unique way to define the speed of a wave packet. At least five velocities can be defined for a wave packet: phase, group, front, signal and energy velocities \cite{brillouin-1960}.
It has been shown experimentally \cite{steinberg-1992} that a single photon with energy far from atomic resonances, travels at the group velocity. 
We determine the group index $n_g$ for the Maxwell-quantum formula through the following relationship: 

\begin{eqnarray}
\label{eqgroupe}
n_g = n_p - \lambda \frac{\delta n_p}{\delta \lambda}
\end{eqnarray}

On the other hand in the corpuscular approach even if there is a statistical distribution of the number of stops, there is a single, definite, average photon velocity. Therefore the corpuscular formula (\ref{eqcorp}) corresponds to the group index and it is to be compared with the group index data.

\subsection{Example of a gas}
\label{sec:airdata}
For many purposes like astronomy, GPS communications and light detection and ranging (LIDAR), the knowledge of the refractive index of air is very important. The number of measurements is large and this is the reason why we choose air as an example of a dilute medium. The group index data are taken from \cite{ciddor-1999}. In the corpuscular model we need to estimate the geometric cross section $\sigma_{\bot}$ of the air molecules. 

Concerning Nitrogen, the mass of $6 \ 10^{23}$ molecules is $28\ g$ and the liquid has a density of $808.6\ kg/m^3$.

Thus a single molecule occupies a volume: $V_{N_2} = 57.7\ 10^{-30}\ m^3$.

The transverse area is thus estimated as $\pi/4\ V_{N_2}^{2/3} =11.7\ 10^{-20} m^2$.

Concerning Oxygen, the mass of $6 \ 10^{23}$ molecules is $32\ g$ and the liquid has a density of $1141\ kg/m^3$.

A single molecule occupies a volume:  $32/(6\ 10^{23}\ 11.41\ 10^5) = 46.7\ 10^{-30}\ m^3$.

The transverse area is thus estimated as   $V_{O_2}^{2/3} = 10.2\ 10^{-20} m^2$.

We calculate then the weighted average transverse area for air :

$< \sigma_{\bot} > = 11.4\ 10^{-20}\ m^2$.

Hence
$$\left[{n_g-1}\right]_{corpuscle}=K{2.55\over 137}10^{25}11.4\ 10^{-20} {197\ 10^{-9} eV m \over{E_{AV}-E_{\gamma}}}=10^{-3}{4.18 K\over{E_{AV}-E_{\gamma}}}$$

In the Maxwell-quantum model we need the number density of the valence electrons. For $N_2$, $n_{valence} = 2\times3$ and for $O_2$, $n_{valence} = 2\times2$. For air this will average to $0.8\times 6+ 0.2\times4 = 5.6$ valence electrons per air molecule. Then under our air conditions we get: $N_{elec} = 14.28\ 10^{25}$ valence electrons$/m^3$. Exactly like what we do for the corpuscular formula we adopt a one term approach with an average energy. The phase refractivity reads:
$$\left[{n_p-1}\right]_{Maxwell}={98.5\ 10^{-3}\over {E_{AV}^2-E_{\gamma}^2}}$$
The energies $E_{AV}$ and $E_{\gamma}$ are in $eV$.
Then we make use of (\ref{eqgroupe}) to get the group index of the Maxwell-quantum approach.

\begin{figure}[htbp]
\centering
\includegraphics[width=12.0cm]{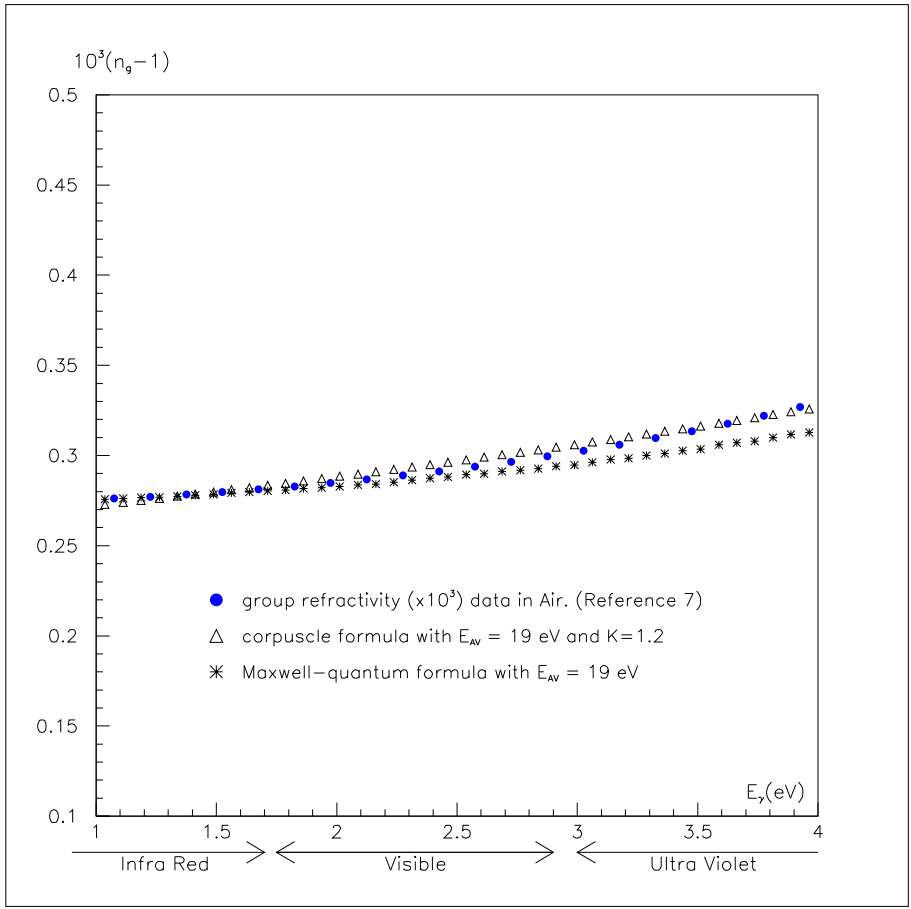} 
\caption{\label{air} {$(n_g-1)10^3$ for air, as a function of the photon energy in $eV$.}}
\end{figure}

The figure \ref {air} shows the Maxwell-quantum and the corpuscle predictions with $E_{AV} = 19\ eV$. The constant $K$ in the corpuscle formula is $1.2$ and this fixes the vertical values. We conclude that the two predictions for air are not very different.

\subsection{Example of a condensed medium}
\label{subsec:SiO2}
As an example of a condensed medium we choose $SiO_2$. The data for synthetic fused silica are in \cite{malitson-1965} and \cite{schneider-2007}. 

The mass of $6\ 10^{23}$ molecules is $60.1 g$ and the density at $25^{\circ} C$ is $2200 kg/m^3$. Therefore, a molecule occupies a volume:  $60.1/(6\ 10^{23}\ 22\ 10^5) = 45.4\ 10^{-30}\ m^3$. 

The spacing is: $\delta_{SiO_2}=(45.4\ 10^{-30})^{1/3}=3.57\ 10^{-10}\ m$, and the corpuscular prediction is:
$$\left[{n_{gSi0_2}-1}\right]_{corpuscle}=K{\alpha\over {\delta_{SiO_2}}} {\hbar c \over{E_{AV}-E_{\gamma}}}=K{4.08\ eV\over{E_{AV}-E_{\gamma}}}$$

In the Maxwell-quantum framework we use the Lorenz-Lorentz formula (\ref{eqLL}).

The number of valence electrons for $SiO_2$ is $4+2\times2 = 8$

The number density is: $2.2\ 10^{28}\times 8 = 17.6\ 10^{28}$ valence electrons $/m^3$.

$$\frac{n_p^2-1}{n_p^2+2}={4\pi\over3}2.83\ 10^{-15}17.6\ 10^{28}3.88\ 10^{-14}\frac{1}{{E_{AV}^2-E_{\gamma}^2}}=A$$
$$\left[{n_{p}-1}\right]_{LL}=\sqrt{1+2A\over1-A}-1$$

Then we calculate $n_g$ from $n_p$ with formula (\ref{eqgroupe}).

\begin{figure}[htb]
\centering
\includegraphics[width=12.0cm]{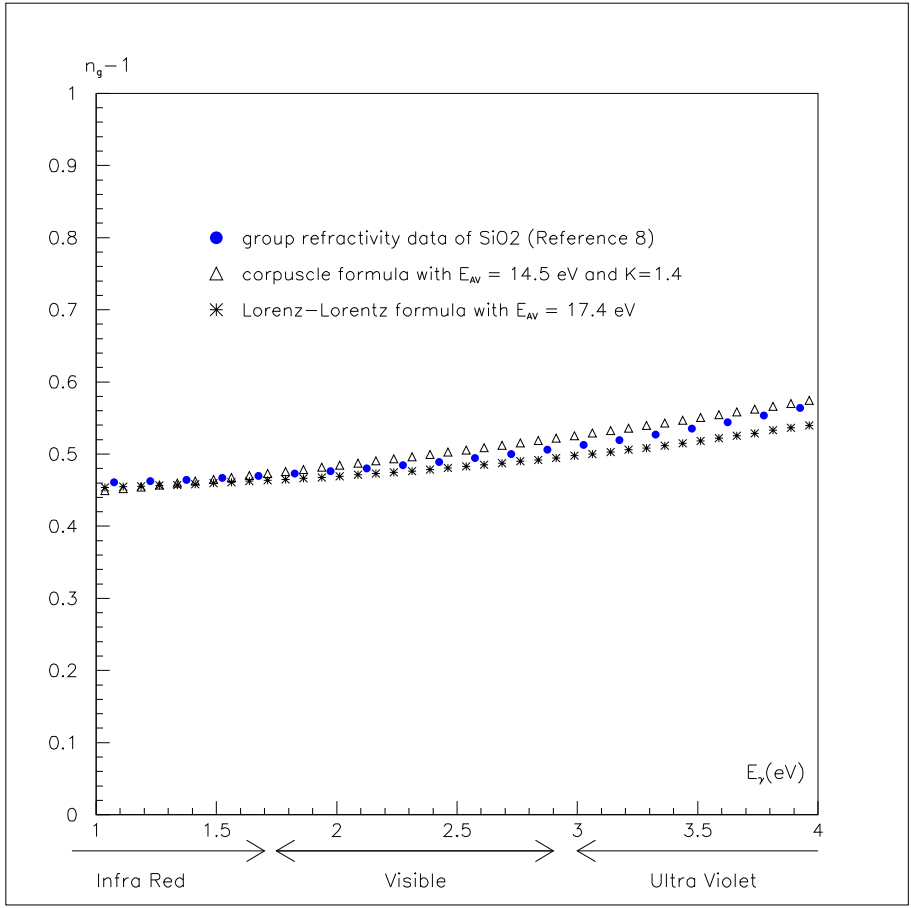} 
\caption{\label{silica} {The fused silica group refractivity: $n_g-1$, as a function of the photon energy in $eV$.}}
\end{figure}

The figure \ref{silica} shows the data together with the corpuscle formula with $K=1.4$ and $E_{AV} = 14.5\ eV$. In order to get a reasonable value for the silica refractivity, we have to plug a value of $17.4 \ eV$ for the average excitation energy in the Lorenz-Lorentz formula. The average excitation energy is lower for the corpuscle formula.

We conclude that the two approaches are very similar when compared to the data either in air or in quartz. The factor K of the corpuscle formula is not too different from 1 as we expected, being 1.2 in air and 1.4 in quartz.

\section{Another perspective for the self focusing Kerr effect}
\label{sec:kerr}
An intense beam of light, focused on a medium, produces filaments several orders of magnitude longer than diffraction would allow. This phenomenon has been observed in solids, liquids and gases and is known by the names of AC Kerr, optical Kerr or self focusing effect. This can be understood if we suppose a major increase of the index of refraction when the light intensity is very high \cite{maker-1964}-\cite{nibbering-1997}. The transverse profile of the laser beam translates into a transverse gradient of the index of refraction. This modified refractive index distribution then acts like a focusing lens allowing the light beam to defeat diffraction. This is usually expressed empirically with the formula:  $n=n_0+n_2I$. $n_0$ is the usual index of refraction at low light intensity $I$. The constant $n_2$ stands around $10^{-20}\ m^2/W$ for dense media and $10^{-23}\ m^2/W$ for gases.

From (\ref{eqsigma}) a photon crossing a molecule has a probability $K_2\alpha$  to stop. If the photon stops and if the incoming flux is high enough to have, at the same time,  $1/K_2\alpha$ real photons on top of the first, we will, on average, get two photons stopping in the molecule. These two real photons will add up their energies to bring the molecule in a virtual excited state. The energy violation will be smaller and therefore the stop time will be larger. This leads to a larger index of refraction. 

Thus when the energy flux of incoming photons $I_{Special}$ corresponds to a density of photons of  $1/K_2 \alpha$ in a molecular volume, the refractivity is multiplied by the factor: $({E_{AV}-E_{\gamma}})/({E_{AV}-2E_{\gamma}})$.

$$(n-1)(I=I_{Special})\approx\frac{{E_{AV}-E_{\gamma}}}{{E_{AV}-2E_{\gamma}}}(n_0-1)$$

\begin{eqnarray}
\label{eqkerr}
n\approx n_0+\frac{E_{\gamma}}{{E_{AV}-2E_{\gamma}}}(n_0-1)=n_0+n_2I_{Special}\Rightarrow n_2=\frac{n_0-1}{I_{Special}}\frac{E_{AV}-2E_{\gamma}}{E_{\gamma}}
\end{eqnarray}

Let us consider visible photons of energy $2.5\ eV$ propagating through water, and estimate the energy flux necessary to get such a special density.
A water molecule occupies approximately, a volume of $30.\ 10^{-30}\ m^3$. 

$$\rho_{photons}={1/(K_2\alpha)\over{V_{molecule}}}={137\over{30.\ 10^{-30}\ K_2}}={4.6\ 10^{30}\over K_2} \ photons/m^3$$
$$\Phi_{photons}=\rho_{photons}{c\over n}={4.6\ 10^{30}\over K_2}{3.\ 10^8\over 1.33}\approx {10^{39}\over K_2}\ photons/m^3/s$$
This can be translated into an energy flux
$$I_{Special}(W/m^2)=\Phi_{photons}E_{\gamma}={2.5\ 10^{20}\over K_2}\ W/m^2$$
$$n_2(water)=\frac{n_0-1}{2.5\ 10^{20}} K_2 \frac{E_{AV}-2E_{\gamma}}{E_{\gamma}}=\frac{0.33}{2.5\ 10^{20}}K_2\frac{17.-5.}{2.5}\approx 0.9\ 10^{-20}\ m^2/W$$
So we understand three things:
\begin{enumerate}\item{}The index of refraction increases when the flux of photons $I$ increases.  $$n=n_0+n_2I$$\item{} $n_2 \approx 10^{-20} m^2/W$ in solids or liquids.
\item{}The refractivity: $n_0-1$, is proportional to the density of the medium. The constant $n_2$ being proportional to $n_0-1$ (\ref{eqkerr}), we understand therefore why $n_2$ is about $1000$ times smaller for gases than for condensed media.
\end{enumerate}

\section{Experimental expectation }
\label{sec:flower}

One prediction of the corpuscular behaviour of photons crossing a homogeneous medium is a fluctuation of the speed of light due to a fluctuation of the number of collisions. The fluctuation would be proportional to the square root of the length of the sample crossed by the photons.

Let's assume photons crossing a fused silica sample of length $L_{SiO_2}$ with a transit time t. The averaged number of collisions $N_{stop}$ is
$$N_{stop} = \frac{L_{SiO_2}}{\Lambda}$$
where $\Lambda$ is the mean free path for a photon in the medium. From equation~\ref{eqtstopdeux}, one obtains
$$ N_{stop} = \frac{1}{t_{stop}} \frac{n-1}{c} L_{SiO_2}. $$ 
We expect a gaussian fluctuation $(\delta_t)_{corpuscular}$ of the transit time of the photons given by
$$ (\delta_t)_{corpuscular} = t_{stop} \times \sqrt{N_{stop}} = \sqrt{t_{stop} \frac{n-1}{c} L_{SiO_2}}$$
From equation~\ref{eqtstopun}, 
one obtains
$$ (\delta_t)_{corpuscular} = \sqrt{\frac{K_1 \hbar (n-1)}{(E_{AV}-E_{\gamma})c}} \times \sqrt{L_{SiO_2}}$$
$E_{AV}=14.5 eV$ has been determined in section ~\ref{subsec:SiO2} from the group index measurement. 
The main uncertainty comes from the value of the constant $K_1$. Assuming a conservative range $K_1=0.1-10$, the transit time fluctuation at $1270 nm$ is expected to be
$$ (\delta_t)_{corpuscular} = \left[150-1500\right] fs \times \sqrt{L_{SiO_2}(m)}$$

The two other main sources of time dispersion are the chromatic dispersion and the quantum dispersion.
The group index of fused silica is almost constant at its minimum value around $\lambda_0 = 1270 nm$ \cite{schneider-2007}. If one apply a spectral filter with a  bandwidth $\delta \lambda = 40 nm$ around $1270 nm$ one would expect a variation of the group index $\delta n_{group} \approx 10^{-5}$ and thus a chromatic dispersion of the transit time 
$$ (\delta_{t})_{chromatic} \approx 35 fs \times L_{SiO_2}(m)$$
A spectral filter of bandwidth $\delta \lambda$ produces also a quantum dispersion. 
From a previous direct measurement of two photon anticoincidence performed in a Hanbury Brown-Twiss type experiment \cite{shih-1994}, with a spectral filter of bandwith $\delta \lambda = 40 nm$, one expects a quantum time dispersion of
$$  (\delta_{t})_{quantum} \approx 30 fs$$

Therefore it appears that the expected fluctuation $(\delta_t)_{corpuscular}$ could be larger than both chromatic and quantum dispersions with a typical length of fused silica of the order of one meter.  
The use of femto laser together with an autocorrelator should allow to measure the fluctuation of the transit time of infrared photons with a precision of $\sim 100 fs$.
The experimental signature is that the expected fluctuation $(\delta_t)_{corpuscular}$ varies like the square root of the optical pathlength $L_{SiO_2}$ crossed by the photons in fused silica when the chromatic dispersion $(\delta_{t})_{chromatic}$ varies linearly and the quantum dispersion is constant.

Let's notice that the measurement of the transit time fluctuation $\delta_t$ would also determine the two constant $K_1, K_2$ of the corpuscular transport of photons, the stop time $t_{stop}$ and the mean free path between collisions $\Lambda$ given by
$$ K_1 \frac{\hbar c}{E-E_{AV}} = t_{stop} = \frac{c}{L_{SiO_2}(n-1)} (\delta_{t})_{corpuscular}^2$$
$$ \frac{1}{K_2 \alpha \sigma_\bot N_{mol}} = \Lambda = \left( \frac{c}{n-1} \right)^2 \frac{(\delta_{t})_{corpuscular}^2}{L_{SiO_2}}$$

\section{Conclusions} 
The bending of light when crossing a boundary between two transparent media can be understood either in terms of wave or in terms of particle. We have given here a coherent description of light velocity reduction in a transparent medium, in terms of particle.
 We do get different, simpler, formula for the index of refraction as compared to the usual Maxwell wave approach and we showed that both reproduce the data fairly well. 

Our corpuscular description leads to new viewpoints and, in particular, we understand and predict the characteristic magnitudes of the self focusing optical Kerr effect both in gases and in condensed matter. 

At last, a measurement of the expected fluctuation of the speed of light seems within the range of today's technical capabilities.

\vskip.2in{\noindent {\bf Acknowledgments}
We wish to thank our colleagues: Barrand G, Haissinski J and Zomer F, for numerous and helpful discussions.
}

\appendix

\renewcommand{\theequation}{A-\arabic{equation}}
\setcounter{equation}{0}

\vskip.2in
{\noindent {\bf Appendix A. The electromagnetic wave point of view for the origin of the index of refraction}


A clear description is made in \cite{feynman-1965}. They consider a continuous slab of dielectric having a small depth and a large extension perpendicular to the axis of propagation of the plane wave. The slab of dielectric is decomposed into small volumes, as compared to $\lambda^3$, where the incident plane wave induces electric dipoles. These elementary vibrating dipoles emit secondary spherical waves. The reason to consider a very thin slab of matter is that the sum of these secondary waves on one particular elementary dipole is negligible compared to the incident light amplitude. The amplitude of oscillation of the elementary dipoles is then proportional only to the amplitude of the incident light. All these secondary waves add up coherently downstream to the incoming light. The coherent integration of these secondary waves is phase shifted by $\pi/2$ with respect to the incident wave because of the integration of the imaginary exponential plane wave. Since the thickness of the slab is small, the amplitude $iA$ of this integration has a smaller modulus than the amplitude of the incident plane wave taken as $1$. The sum, downstream, of the incident light and of the secondary waves takes the form $1+iA$ which is approximated as $e^{iA}$. Thus instead of being 1 the incident wave is transformed into $e^{iA}$ which displays a phase shift interpreted as a delay in time due to a smaller speed in the slab of dielectric. This at last gives the index of refraction.
The strength of the induced electric dipoles is given by their amount of polarization under the influence of the incident electric field. This approach was envisaged for a continuous distribution of matter, and when the atomic nature of matter was established it became possible to have a microscopic model for the induced dipoles. The electric field of the incident wave sets the electrons in the molecule into a dipolar periodic motion and these electrons are supposed to reemit light spherically and at the same frequency as the one of the incoming wave. 
Then we have to go from a discrete to a continuum, which as shown in \cite{rosenfeld-1951} is not so obvious. The continuum is necessary in order to use the framework of the coherent integration producing the phase shift and thus explaining the index of refraction.
Any approach to the prediction of the index of refraction is based upon the calculation of the average electric polarization of the molecule, produced by the incident electric field of the plane wave. Then go to the macroscopic polarization. Once in the continuum theory framework the dielectric constant $\epsilon$ is obtained as a function of frequency and finally the index of refraction is predicted through the Maxwell formula: $n^2=\epsilon$. We call this line of thought the Maxwell-quantum framework. The quantum predictions of the index of refraction can be found in textbooks \cite{rosenfeld-1951} and \cite{lorentz-1952}. \begin{eqnarray}
\label{eqfi} 
n^2-1=4\pi r_e N_{elec} ({\hbar c})^2{\sum_{j}{{f_j}\over {(W_j-W_0)^2-E_{\gamma}^2-2i\delta_jE_{\gamma}}}} \ \ \ ,\ \ \ {\sum_{j}{f_j}}=1
\end{eqnarray}

$\delta_j$ are the damping constants of the levels. When the life time of the state $j$ is large, the energy $\delta_j$ is small. The ground state energy of the molecule is $W_0$ and the excited states have energies $W_j$. $N_{elec}$ is the number of valence electrons per unit volume and $r_e = 2.83\ 10^{-15} \ m$, is the classical radius of the electron. ${\hbar c}=197.3\ MeVfm$. $f_i$ is the relative strength of the transition $W_0 \to W_i$. The energy of the photon is $E_{\gamma}$.
The Hydrogen atom has $W_1-W_0 = -3.4+13.6 = 10.2\ eV$. And the life time of the $Ly\alpha$ corresponds to $\delta_1=2.\ 10^{-6}\ eV$. The damping terms are therefore neglected in this paper.
The discrete sum in (\ref {eqfi}) is always replaced by a single term (sometimes two) with an average excitation energy. It came as a surprise, at the beginning of the $20^{th}$ century that this average energy was much greater than the typical molecular excitation energies. It was realized then that along with the discrete energies, the ionization continuum should be included thus explaining why the average energy was higher than expected. This ionization term is large. In Helium it is between 2.6 and 3 times as large as the discrete sum term \cite{vinti-1932} and \cite{wheeler-1933}.
In a condensed medium, Lorentz and Lorenz try to take into account the difference between the incident and the local electric field and suggest the following formula:
\begin{eqnarray}
\label{eqLL}
\frac{n^2-1}{n^2+2}={4\pi\over3}r_e N_{elec} ({\hbar c})^2{\sum_{i}{{f_i}\over {(W_i-W_0)^2-E_{\gamma}^2}}}
\end{eqnarray}

\vskip.2in
{\noindent {\bf Appendix B. The cross-section for the absorption of a photon far from the resonance.}

\renewcommand{\theequation}{B-\arabic{equation}}
\setcounter{equation}{0}

The cross section for the absorption of a real photon by an atom can be calculated in the framework of quantum mechanics and perturbative theory (assuming a small perturbation).
The photon induces transitions between the atom's states $|k>$.
Assuming a first order transition for the absorption process, the transition probability reads~\cite{sakurai-1985}\cite{cohentann-1992}
\begin{eqnarray}
\label{eq:TransitionProbability}
P_{i\rightarrow k}(\tau)=\left| \frac{1}{i\hbar}\int_{0}^{\tau}dt <k|H_1(t)|i> e^{i(E_k-E_i)t/\hbar}\right|^2 .
\end{eqnarray}
$\tau$ is the duration of the perturbation.
The perturbation Hamiltonian $H_1(t)$ can be defined by the product of a time shape perturbation function
$f(t)$ (with $f(t)=0$ for $t<0$ and  $t>\tau$) with a time independent operator $\widehat{W}$ acting on initial state $|i>$ and the final state $|k>$ and an exponential factor $e^{-i\omega t}$ corresponding to the time dependence of the incident photon wave.
The transition probability can be written as~:
\begin{eqnarray}
P_{i\rightarrow k}(\tau)=\frac{2\pi}{\hbar}\left|<k|\widehat{W}|i>\right|^2 \tau \delta_\tau(E_k-E_i- \hbar\omega) 
\end{eqnarray}
where $\delta_\tau(E)$ is a zero-peaked function, of width $2\pi\hbar/\tau$, of height $\tau/(2\pi\hbar)$, of unit integral such that $\lim_{\tau\rightarrow \infty} \delta_\tau(E)=\delta(E)$ (the Dirac distribution). 
For $f(t)$ taken to be a squared window shape of width $\tau$, 
\begin{eqnarray}
\delta_\tau(E)=\frac{\tau}{2\pi\hbar}\frac{\sin^2{\frac{\tau E}{2\hbar}}}{\left(\frac{\tau E}{2\hbar}\right)^2}
\end{eqnarray}
The time $\frac{2\hbar}{E}$ can be larger than the crossing time of the molecule as we saw in section~\ref{sec:propagas}. Therefore we will approximate $\delta_\tau$ as $\delta_\tau=\frac{\tau}{2\pi\hbar}$.

The interaction at the lowest order between an electromagnetic wave and the electrons of an atom is the dipolar interaction given by $\widehat{W}=-q\widehat{r}{\cal E}$ where $q$ is the electron charge,  ${\cal E}$ is the electric field induced by the photons, 
$\widehat{r}$ is the operator of the radial distance between the electron and the nuclei.
The magnitude of the electric field is related to the photon density by $\epsilon_0 {\cal E}^2 = n_\gamma \hbar \omega$.
For a given photon flux $n_\gamma c$, the cross-section can be calculated by
\begin{eqnarray}
\sigma_{ik}(\tau)=\frac{P_{i\rightarrow k}(\tau)}{n_\gamma c \tau}
\end{eqnarray}






The matrix element of the Hydrogen $|<2p|z|1s>|^2=\left(\frac{2^7\sqrt{2}}{3^5}a_0\right)^2$, where $a_0$ is the Bohr radius. The geometrical cross section for the Hydrogen is thus $\sigma_\bot=\pi a_0^2$ and we can write $|<2p|z|1s>|^2\approx 0.16 \sigma_\bot$.

We end up with $\sigma_{ik}(\tau)\approx 2 \alpha \sigma_\bot \omega\tau$. From Heisenberg relation, we expect $\omega \tau \approx 1$, thus  $\sigma_{ik}(\tau)\approx 2 \alpha \sigma_\bot$, to be compared to our guess $\sigma_{ik}(\tau)\approx K \alpha \sigma_\bot$ in section~\ref{sec:propagas}.

\end{document}